\documentstyle [ emulateapj]{article}

 1

\def\mpc{\,{\rm {Mpc}}}

\def\kms{\,{\rm {km\, s^{-1}}}}
\def\msun{{M_\odot}}

\def\etal{{et al.~}}
\def\HH{${\rm {H_2}}\,\,$}
\def\erg{{\rm erg}}
\def\sr{{\rm sr}}

\def\cm{{\rm cm}}
\def\nhi{{N_{\rm HI}}}

\def\gs{\mathrel{\raise1.16pt\hbox{$>$}\kern-7.0pt
\lower3.06pt\hbox{{$\scriptstyle \sim$}}}}
\def\ls{\mathrel{\raise1.16pt\hbox{$<$}\kern-7.0pt
\lower3.06pt\hbox{{$\scriptstyle \sim$}}}}
\def\gtsima{$\; \buildrel > \over \sim \;$}
\def\ltsima{$\; \buildrel < \over \sim \;$}
\def\prosima{$\; \buildrel \propto \over \sim \;$}
\def\gsim{\lower.7ex\hbox{\gtsima}}
\def\lsim{\lower.7ex\hbox{\ltsima}}
\def\simgt{\lower.7ex\hbox{\gtsima}}
\def\simlt{\lower.7ex\hbox{\ltsima}}
\def\simpr{\lower.7ex\hbox{\prosima}}
\def\la{\lsim}
\def\ga{\gsim}

\def\pp{\noindent\parshape 2 0truecm 17truecm 2truecm 15truecm}
\def\rf#1;#2;#3;#4 {\par\pp#1, #2, #3, #4. \par}

\def\pr{\ref@jnl{Phys.Rev}}

\def\href#1;#2 {{\bf #1} : {\em #2}}

\def\beq#1{\begin{equation}\label{#1}}
\def\eeq{\end{equation}}
\def\beqa#1{\begin{eqnarray}\label{#1}}
\def\eeqa{\end{eqnarray}}
\def\eq#1{equation~(\ref{#1})}

\def\tento#1{\times 10^{#1}}

\def\K{{\rm \ K}}
\def\s{{\rm \ s}}
\def\sr{{\rm \ sr}}

\def\erg{{\rm \ erg}}
\def\cm{{\rm \ cm}}
\def\eV{{\rm \ eV}}

\def\pc{{\rm \ pc}}

\def\Hz{{\rm \ Hz}}

\def\nhi{N_{\rm HI}}
\def\HH{H$_2$ }
\def\H2p{H$_2^+$ }

\def\mH2p{H_2^+}

\makeatletter

\newenvironment{figurehere}
  {\def\@captype{figure}}
  {}
\makeatother

\begin{document}
\thispagestyle{empty}
\textheight=25cm

\title {    A ``Minihalo'' Model for the Lyman Limit Absorption Systems \\
                          at High Redshift}
\author {Tom Abel\altaffilmark{1,2} and  H. J. Mo\altaffilmark{2}}
\received{29 August 1997}
\accepted{ApJ Letters, 8 Dec 1997 }

\altaffiltext{1}{ Laboratory for Computational Astrophysics, NCSA,
          University of Illinois at Urbana/Champaign, 405 N. Mathews
          Ave., Urbana, IL 61801.}  \altaffiltext{2}{
          Max-Planck-Institut f\"ur Astrophysik,
          Karl-Schwarzschild-Strasse 1, 85748 Garching, Germany\\
	{\bf ApJ Letters accepted } }

\begin{abstract}
  We propose that a large fraction of the QSO Lyman limit absorption
  systems (LLS) observed at high redshift ($z\ga 3$) originate from
  gas trapped in small objects, such as minihalos, that form prior to
  reionization.  In the absence of a strong UV flux, the gas is
  predominantly neutral and may form clouds with HI column density
  $\nhi\gsim 10^{18}\cm ^{-2}$.  Due to their high densities and high
  HI column densities, these clouds are not destroyed by the onset of
  the UV background at a later time.  Thus, if not disrupted by other
  processes, such as mergers into larger systems or `blow away' by
  supernovae, they will produce LLS.  We show that the observed number
  density of LLS at high redshifts can well be reproduced by the
  survived `minihalos' in hierarchical clustering models such as the
  standard cold dark matter model.  The number density of LLS in such
  a population increases with $z$ even beyond the redshifts accessible
  to current observations and dies off quickly at $z\la 2$. This
  population is distinct from other populations because the absorbing
  systems have small velocity widths and a close to primordial
  chemical composition. The existence of such a population requires
  that the reionization of the universe occurs late, at $z\lsim 20$.
\end{abstract}
\setcounter{page}{1}
\section{Introduction}

 Lyman limit absorption  systems (LLS), observed  as breaks in  quasar
spectra at the rest-frame wavelength  of the Lyman continuum edge, are
produced  by gaseous    clouds optically  thick  to Lyman    continuum
photons. These systems can be observed  over a wide range of redshift,
from $z\sim 0$ to approximately that  of the highest-redshift quasars.
Thus, such systems  can be used  to probe the  gas in  the Universe at
various  redshifts.   Due to their  relatively  high HI column density
($\nhi \gsim 10^{17}\cm ^{-2}$),  these systems are usually thought to
arise from photoionized  clouds   in   gaseous galactic halos     (see
e.g. Steidel 1990, Mo \& Miralda-Escud\'e 1996).   There is now direct
observational evidence that some of the LLS are indeed associated with
galaxies (see e.g. Steidel 1995), at least at redshift $z<3$.  This is
not surprising, because  any HI disk like that  of our own Galaxy will
produce a LLS when intersecting a sightline to a quasar.  However, the
observed rate of incidence (i.e.  the number  of systems per sightline
per unit redshift) of LLS is large and increases rapidly with redshift
(Stengler-Larrea \etal 1995); there is no  indication of a reversal of
this trend even   for the LLS with  the  highest redshift  $z\sim 4.5$
observed so far (Storrie-Lombardi \etal  1994).  The observed rate  of
incidence  of  LLS at  $z\sim 4$ is   about 30 times  larger than that
expected from  the local number  density  of galaxies with  absorption
cross sections given by the optical  Holmberg radius.  Thus, if LLS at
$z\gsim 4$ were also associated with  galaxies, it would mean that (1)
most  galaxies have already   formed  at that  redshift,  and (2)  the
absorption  cross sections of     galaxies  are much larger    at high
redshifts.  However, in a hierarchical  model of structure  formation,
such as the  standard cold dark  matter (CDM) model  and its variants,
the number density and sizes of  galaxies are small at high redshifts,
because most gas has not yet been assembled into large galaxies (Mo \&
Miralda-Escud\'e 1996).  It is therefore unlikely that LLS observed at
high redshifts are due to galactic populations.

  It is possible that a large fraction of the observed LLS at high
redshifts are produced by photoionized clouds in small halos.  Recent
hydrodynamic simulations show that low column density absorption
systems can be produced by gas in filamentary structures that arise
from the gravitational collapse of initial density fluctuations (e.g.
Cen \etal 1994; Zhang, Anninos, \& Norman 1995; Hernquist \etal 1996;
Haehnelt, Steinmetz \& Rauch 1996; Miralda-Escud\'e \etal 1996).  The
number densities of LLS given by these simulations were, however,
found to be lower by a factor of about ten than what is observed (e.g.
Katz \etal 1996). The discrepancy at $z\la 2$ may be explained by the
population of pressure-confined clouds in gaseous halos that are not
resolved by current simulations (Mo \& Miralda-Escud\'e 1996).  The
discrepancy at $z\ga 3$, however, was not explained by previous
investigations based on photoionization models in hierarchical
cosmogonies (e.g. Gardner \etal 1996). 

 It is also possible that the
flux of the UV background may drop substantially at $z> 3$ and the HI
column density of a cloud could be underestimated in models assuming a
constant UV flux. However, recent observational results based on the
proximity effect show that the UV flux at the Lyman continuum edge is
about $(1-5)\times 10^{-22} {\rm erg\, s^{-1}\, \cm^{-2}\,Hz^{-1}\,sr
^{-1}}$ at $z\sim 4.5$ (Williger \etal 1994; Giallongo \etal 1996; Lu
\etal 1996). This value of the UV flux is already comparable to, or
even higher than, the ones used in most of the model considerations
mentioned above.

  In this paper we suggest the possibility 
that the discrepancy between model
predictions and observations of the LLS at high redshifts ($z\gsim3$)
is a result of the neglect of the contribution of minihalos that
collapsed prior to reionization.  The gas in such small systems is
predominantly neutral and may form clouds with \ion{H}{1} column
density $\nhi\gsim 10^{18}\cm ^{-2}$.  Because of their high density
and high HI column density, these clouds will not be destroyed, or
ionized by the onset of the UV background at a later time.  They will
therefore produce LLS at lower redshifts, if not disrupted by other
processes. 

We present our model for the LLS observed at high redshifts and
compute their number per unit redshift in Section~2.  A brief
discussion of our model is given in Section~3.

\section{A New Population of Lyman Limit Systems} 

For simplicity we assume  the  Universe  to be  flat with  the  cosmic
density parameter  $\Omega_0=1$.   The Hubble  constant  is written as
$H_0=100h\kms \mpc^{-1}$.
\subsection{Physical properties}
 Prior  to   the epoch  of   reionization  the physical properties  of
cosmological structures can  be estimated from the spherical  collapse
model (see e.g. Padmanabhan,  1993).   The gas in a
collapsing dark  matter perturbation  is  shock  heated to the  virial
temperature
\begin{eqnarray}
  \label{f} T_{vir} \equiv \frac{\mu V_c^2}{3k}\approx 4.0\tento{3}\K
  \left( \frac{V_c}{10\kms} \right)^2 \left( \frac{\mu}{m_P} \right) ,
\end{eqnarray}
where  $V_c$ is  the three  dimensional  virial velocity of  the halo,
$m_P$ the proton mass, and $\mu$ the mean mass per particle.

The total H number  density in a dark  matter halo that collapses at a
redshift, $z$, with an overdensity $\delta$ at virialization reads
\begin{eqnarray}
  \label{nvir} n_{vir} & = &  
  \frac{\Omega_B \rho_{\rm crit}\delta }{m_P} (1+z)^3 
  \approx  9.7\tento{-3}  \nonumber \\
   &\times & \left( \frac{\Omega_B
   h^2}{0.0125} \right) \left( \frac{1+z}{10} \right)^3 \left(
   \frac{\delta}{18 \pi^2} \right) \cm^{-3},
\end{eqnarray}
where $\rho_{crit}$ denotes the  closure density.  Assuming  that dark
halos   are   singular   isothermal  spheres  with    density  profile
$\rho(r)\propto r^{-2}$, the    virial radius $r_v$, defined   as  the
radius  within  which the mean   mass overdensity is~$18\pi^2$, can be
written:
\begin{eqnarray}
  \label{rvir} 
  r_{vir} = \frac{V_c t}{2 \pi} \approx
  350 \pc \ h^{-1} \frac{V_c}{10 \kms} \left( \frac{1+z}{10} \right)^{-3/2},
\end{eqnarray}
where $t=(2/3 H_0) (1+z)^{-3/2}$ is the cosmic time at $z$.  

For halos with $T_{vir} \lsim 10^4\K$ ($V_c \lsim 15\kms$) the gaseous
component cannot be collisonally ionized and will remain neutral. 
Due to the high collapse redshifts ($z>5$) considered here, the
recombination time scale in minihalos is shorter than the Hubble
time, and the gas in clouds with $V_c \gsim 15\kms$ before reionization
will also be neutral shortly after their formation.  As a result, the column
density of neutral hydrogen atoms, $\nhi$, is given by
\begin{eqnarray}
  \label{NHI}
  \nhi & \approx & 2 r_{vir} n_{vir} \approx 2.1 \tento{19} 
  \left( \frac{V_c}{10\kms}  \right) \nonumber \\ 
  &\times &
 \left( \frac{1+z}{10} \right)^{3/2}  
  \left(  \frac{\delta}{18 \pi^2} \right)
  \left( \frac{\Omega_B h^2}{0.0125} \right)  
   \cm^{-2}.
\end{eqnarray}
The corresponding total mass of such a halo is
\begin{eqnarray}
  \label{mass}
  M = \frac{V_c^3 t}{2\pi G} 
	&\approx& 7.9\times 10^6 \left({V_c\over 10\kms} \right)^3 
     \nonumber \\ 
   & \times&
	 \left({1+z\over 10}\right)^{-3/2} \,h^{-1}\msun .
\end{eqnarray}
It is clear that the optical depth at the photoionization threshold,
\begin{eqnarray}
 	\tau = \sigma(\nu_{\rm TH}) \nhi &\approx& 130 
 	 \left(
	  \frac{V_c}{10\kms} \right) \left( \frac{1+z}{10}
	  \right)^{3/2}
	 \nonumber \\ 
 &\times &
	 \left( \frac{\delta}{18 \pi^2} \right) \left(
	  \frac{\Omega_B h^2}{0.0125} \right) \cm^{-2},
\end{eqnarray}
exceeds unity for a wide range of collapse redshifts and virial
velocities. 

To determine   what  happens to  the  clouds  once  the  UV background
switches  on  we analyze   the  equilibrium equation  of   the ionized
fraction for a pure hydrogen gas, $x\equiv n_e/n$, where $n_e$ and $n$
denote   the  free  electron and     total  baryonic  number  density,
respectively. This fraction is given by
\begin{eqnarray}\label{equ1}
0 =  k_{Ph} n(1-x) - k_{Rec} n^2x^2,
\end{eqnarray}
where  $k_{Ph}=4  \pi \int_{\nu_{  TH}}^\infty   (J/h \nu) \exp[- \nhi
\sigma(\nu)] \sigma    (\nu)   d\nu$ and $k_{Rec}$   denote   the rate
coefficients  for photoionization of   neutral hydrogen  and radiative
recombination to  \ion{H}{1}, respectively.  Motivated by the indirect
measurements from the proximity effect  (see Giallongo \etal 1996, and
references therein) we assume the background UV flux to  be given by a
powerlaw
\begin{eqnarray}\label{UV}
  J(\nu) = 10^{-21} J_{21}  
  \left( \frac{h\nu}{1 \rm Ryd} \right)^{-\alpha}  
  \erg \cm^{-2}  (\Hz \s \sr) ^{-1},
\end{eqnarray}
where $h$ is Planck's constant and $h\nu_{TH}=1\rm Ryd \equiv 13.6\eV$
is the \ion{H}{1} ionization threshold. In the optical thick limit the
photoionization  rate coefficient can  be  approximated by a power law
$k_{Ph} \approx J_{21} k_{Ph}^0 [N_{H}(1-x)]^{-\beta}$, where $N_H$ is
the total column density of hydrogen  nuclei. The exponent $\beta$ has
a non--trivial dependence on $\alpha$ and is determined numerically to
vary from  $1.35$  to $1.69$   for $1<  \alpha  <2$.  Inserting   this
approximate   expression for   the   photoionization   rate  into  the
equilibrium equation for the free electron fraction [\eq{equ1}] yields
the implicit solution for~$x$
\begin{equation}\label{equ2}
   f(x) \equiv  \frac{x^2}{(1-x)^{1-\beta}} = 
      \frac{J_{21}k_{Ph}^0N_H^{-\beta}}{n k_{Rec}} \equiv C,
\end{equation}
which for $\beta=0$ reduces to the more familiar optical thin case.
We adopt $\alpha = 1.8$ for the spectral index, with which the
photoionization rate can be approximated by $k_{Ph} \approx
4.1\tento{-15} J_{21} (\nhi/10^{19}\cm^{-2})^{-\beta} \s^{-1}$ with
$\beta=1.6$.

The right hand  side of \eq{equ2} has  a maximum at $x=x_{max} \approx
0.8 $  with a value of $f(x_{max})   \approx 0.25$.  The  existence of
this maximum shows that ionization equilibrium  cannot be achieved for
$J_{21}k_{Ph}^0N_H^{-\beta}/(n k_{Rec})  > f(x_{max}) $.  In this case
photoionization overcomes recombination and   ionizes the gas until  a
new, highly ionized, equilibrium  state is reached at column densities
$\ll  10^{17} \cm^{-2}$.  As  discussed in M\"ucket  and Kates (1997),
the  evolution of such  systems from one  equilibrium state to another
can be rapid.  The recombination  rate\footnote{Note  that the Case  B
recombination rate is    used for the  optical   thick case considered
here.}    suitable  for  an  equilibrium  temperature  of  $10^4\K$ is
$k_{Rec}\approx 2.6\tento{-13}  \cm^3  \s^{-1}$  (Ferland \etal 1992).
Thus, for a  given  $J_{21}$, the  maximum  of $f(x)$  can  be used to
derive a  critical total hydrogen  column density $N_{H}^{crit}$. Note
that this column density is calculated at the edge of the cloud and is
approximately equal to $r_{vir}n_{vir}$ [see \eq{nvir} and \eq{rvir}].
We can therefore  define a minimum  collapse redshift, $z_{crit}$, for
halos of  a given virial velocity, so  that they will remain optically
thick if they form at $z>z_{crit}$. This minimum collapse redshift can
be written as
\begin{eqnarray}\label{zcrit}
 1+z_{crit}   &\approx&   7.2 
       \left(  \frac{J_{21} }{0.5 }     	\right)^{5/27}
       \left(  \frac{V_c }{10\kms} 	\right)^{-8/27 }
       \nonumber \\ &\times &
       \left(  \frac{\delta }{18 \pi^2 }      	
       	       \frac{\Omega_B }{0.05 }      	\right)^{-13/27 }
       \left(  \frac{h }{0.5 }     		\right)^{-2/3 }.
\end{eqnarray}
As we will show below, minihalos (with $V_c\sim 10\kms$) in current
hierarchical models have typical collapse redshifts higher than this
critical value; they can therefore retain their high initial
\ion{H}{1} column densities.

To check the  robustness   of the above  derivation for   the critical
collapse redshift we have integrated  the time-dependent chemistry and
cooling model presented in Abel \etal~(1997a). We found good agreement
between   the numerical results  and the   analytic expression of  the
critical redshift given by \eq{zcrit}. 

The  density  of gas contained in  minihalos  considered here is quite
high  due  to their high collapse    redshifts. The recombination time
scale for the gas  is short, about $0.1~$Myr,  which makes them stable
against photoionization from  internal stellar sources.   On the other
hand, only a few supernova explosions may be able  to blow out most of
the gas  from the halos (see Ciardi  \&  Ferrara, 1997, and references
therein) if star formation can happen in  them.  However, systems that
form in halos with $V_c\lsim 15\kms$ at  $z\lsim15$ are not be able to
cool  since \HH  formation is   inefficient  in  these  halos at  such
redshifts (see Tegmark \etal 1997, and Abel \etal 1997b).  Also the UV
flux of  the  first  structures  in  the universe   may  lead to   \HH
photodissociation prior  to \ion{H}{1} reionization (Haiman,  Rees and
Loeb~1997), inhibiting line   cooling by \HH.  Furthermore  \ion{H}{1}
line cooling is not efficient due to the low virial temperature ($\sim
10^4\K$). Thus  these  systems may not form  stars  and hence  are not
subject to ``blow away''.
The gas can  only be gravitationally confined  in the dark matter halo
if the  typical particle velocity is less  than the escape velocity of
the  halo.   The  mean  particle  velocity  of  hydrogen  atoms   in a
Maxwellian distribution  is  $V_H(T)= (3kT/m_P)^{1/2}$  for  a neutral
gas,  and  the escape velocity   is given  roughly by  $V_{esc}\approx
\sqrt{3} V_c$. Hence only systems with $V_c \gsim V_H(10^4\K)/\sqrt{3}
\approx 9 \kms$ are stable against evaporation. These considerations
suggest that minihalos with $V_c=10$-$20\kms$ are most likely to 
produce LLS. 

\vspace{-0.2cm}
\begin{figurehere}
\epsscale{0.5} \plotone{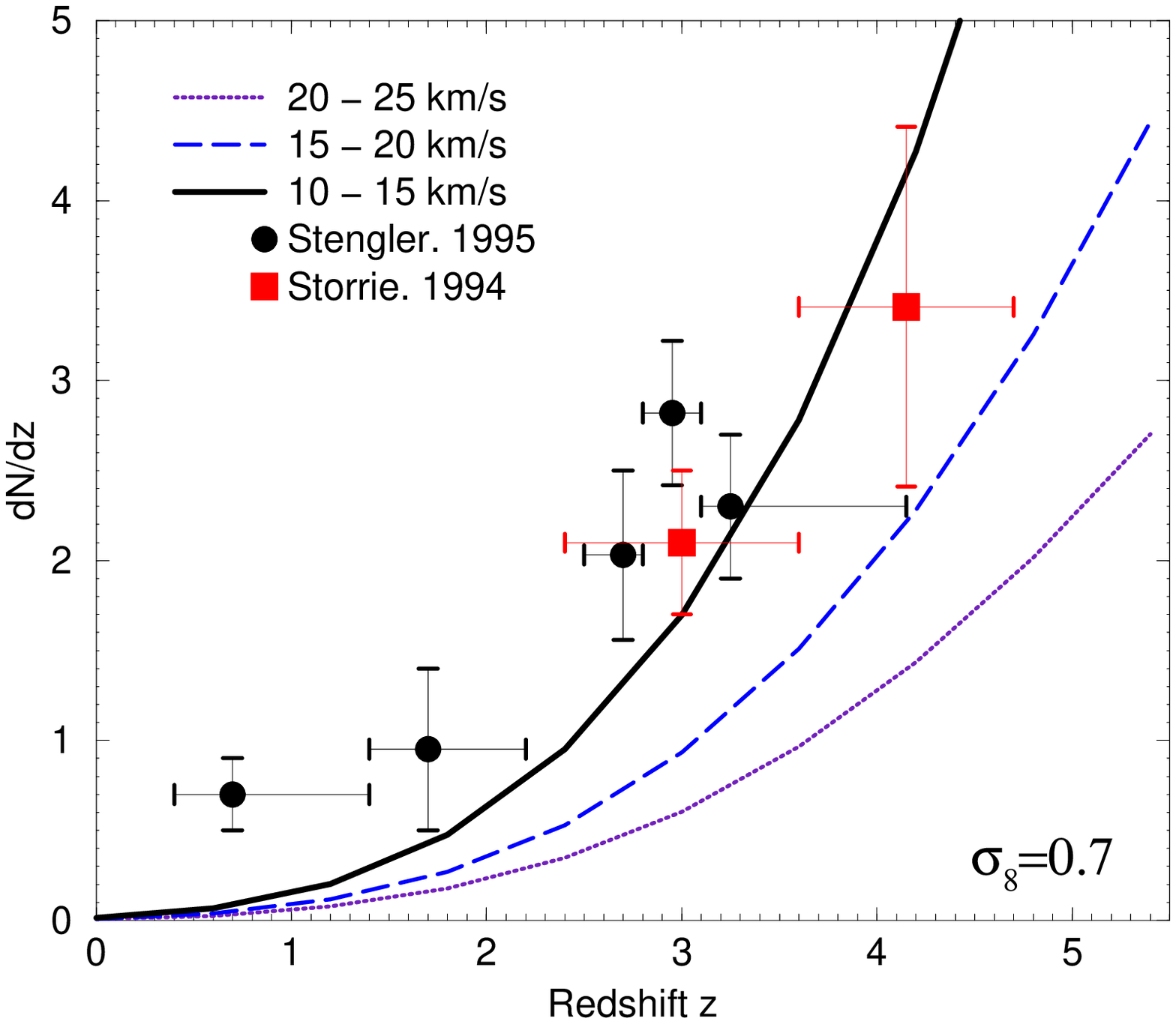}\vspace{-0.8cm}
\caption{ \footnotesize The number of LLS per unit redshift.  The
curves show the prediction of the Standard CDM model with
$\sigma_8=0.7$.  Results are given for minihalos in three $V_c$ bins.
The proposed minihalo LLSs provide a significant fraction of all LLS
only at redshifts $z\simgt2$. The observational data is taken from
Stengler-Larrea \etal (1995) (filled circles) and Storrie--Lombardi
\etal (1994) (filled squares).  }\label{n(z)}\vspace{0.1cm}
\end{figurehere}

\subsection{The predicted number density of LLS}

The typical   collapse redshift of   halos with  $V_c\lsim 30\kms$  in
current  cosmogonies such as    the standard  CDM model exceeds    the
critical  redshift  given in \eq{zcrit}.   Such   small halos are also
abundant at  high redshift in  these models.  It is therefore possible
that  minihalos contribute a substantial part  of  the total number of
LLS at high redshifts.  Here we examine this possibility.

We assume that minihalos which  form at $z>z_{crit}$  and do not merge
into  larger systems   by a redshift   $z_s$  will contribute to   the
absorption cross section    at $z_s$.   The  conditional  probability,
$p(M_2,z_2|M_1,z_1)dM_2$, that   a  halo of   mass $M_1$  selected  at
redshift $z_1$ will have merged  to form a halo  of mass between $M_2$
and  $M_2+dM_2$ at a later redshift  $z_2$, can be calculated from the
extended Press-Schechter formalism (Bond \etal 1991;  Bower 1991).  As
in Lacey \& Cole (1994), halos are assumed to be destroyed by redshift
$z_s$, if they have  merged into halos of twice  their initial mass by
that  redshift.  The  fraction  of minihalos  that  form  at $z_1$ and
survive merging until $z_s$, $f_s$, can then be obtained from
\begin{eqnarray}
f_s(M_1, z_1, z_s)  = 1 - \int_{2M_1}^\infty p(M_2,z_s|M_1,z_1)dM_2.
\end{eqnarray}
\noindent
The absorption cross section of a minihalo is assumed to be spherical
with radius given by \eq{rvir}.  The result for a standard CDM model
normalized to $\sigma_8=0.7$ is given for three different bins of virial
velocities in Figure~1.  In the calculation  $z_1$ is chosen as the
redshift at which  the comoving number density of halos prediced by the
Press-Schechter formalism peaks. This corresponds to the collapse
redshift of a $\sim 1.5\sigma$ peak. In all cases, $z_1>z_{crit}$. 

\figurehere Clearly, the total number of systems with $10\kms \lsim
V_c \lsim 15 \kms$ is sufficient to explain the observed number of
LLSs at $z\ga 3$.  This is consistent with the observational evidence
that LLS at low redshift are associated with galaxies (see Steidel
\etal 1996).  As shown in Mo \& Miralda-Escud\'e (1996), clouds
pressure-confined in galactic halos can indeed give a sufficiently
large cross section to explain the number density of LLS at $z\la 2$.
Note that for current low-$\Omega$ cosmogonic models 
having smaller power on small scales than the standard CDM model, the 
number density of minihalos may be reduced. However, the 
predicted number of absorption systems per unit redshift
may not be reduced substantially, because it
depends also on $dl/dz$ (the proper distance per unit redshift)
and the virial radius, both are larger in a low-$\Omega$ model.   
The dependence on cosmological parameters cannot be examined
in detail until more detailed hydrodynamical calculations are able to
determine the physical sizes of the absorbers.

It is important to realize the  difference  between our model and the
minihalo   model proposed   by Rees (1986)     for Lyman alpha  forest
systems. In  the model of Rees, minihalos   are optically thin, highly
ionized and have    an HI column density  of   $\ll 10^{17}\cm ^{-2}$,
characteristic for  Lyman  alpha forest  systems.   This assumption is
correct for minihalos that collapse after reionization from an already
ionized     intergalactic     medium    at redshifts      $z<z_{crit}$
[see~\eq{zcrit}].

\section{Discussion}

As shown in Thoul and Weinberg (1996), after reionization the collapse
of gas  can  only happen  in  halos   with $V_c\gsim30\kms$,  and   so
minihalos (with $V_c\lsim30\kms$) that form after reionization are not
expected to give a significant contribution  to the LLS.  In contrast,
prior to  reionization the  collapse of  gas can  even happen in halos
with  $V_c$ as small as about  $0.1 \kms (1+z_{coll})$.  Our model for
the LLS envisages  a hierarchical  cosmogony where  a large  number of
minihalos  (with $V_c \lsim 30 \kms$)  can form prior  to the epoch of
reionization.  We show that the gas trapped  in such minihalos is, due
to  its high  density, able  to withstand   photoionization by the  UV
background  set on at  a later time and  reaches  thermal and chemical
equilibrium at   a  temperature $\sim  10^4\K$.   By  including  these
systems,   the discrepancy  in    the   number of  LLS  between    the
observational result and earlier  theoretical predictions based on CDM
models  can be reconciled.  The model  predicts  the number density of
LLS to keep increasing towards higher redshifts.

 The minihalo population is assumed to form at a  time when the gas in
the IGM is largely primordial. As a result, the absorbing gas in these
systems should also have a close to  primordial composition.  Although
later    accretion of  gas  from  the  enriched  IGM  may increase the
metallicity, we expect the metallicity of such LLS to be low.

 Minihalos  with $10  \kms \lsim V_c  \lsim 15  \kms$  are expected to
dominate the total cross section of LLS at high redshifts, because (i)
they  can form  in the low-pressure IGM prior  to
reionization, (2) they are  stable against evaporation, (3) they cannot
cool by \HH or \ion{H}{1} to  form   stars and so are   not
subject to supernova explosion   or internal photoionization, and  (4)
their total cross section is large  in hierarchical models.  Thus, the
minihalo population of  the LLS should  be observed  to  have a narrow
range of velocity widths (with \ion{H}{1} Doppler parameters of $\lsim
20\kms$).  This can be tested  by high-resolution spectroscopy of  LLS
at high redshifts. The two  LLS at redshifts 3.32  and 2.80 studied by
Tytler \etal (1996) using the HIRES spectrograph on the KECK telescope
have Doppler   parameters and  velocity  offsets   between  individual
components  less    than  $20\kms$. This   is   consistent  with   our
model.  Obviously   more high-resolution data   at  high redshifts are
needed  to constrain the fraction of   LLS in the minihalo population.
The sizes for the minihalo absorbers are  typically about 1 kpc, which
may be tested by the spectra of gravitationally-lensed quasar pairs at
high redshifts.

Much theoretical work remains to be done to model the hydrodynamic and
radiative-transfer processes to obtain more accurate predictions for
the properties of the minihalo Lyman limit systems proposed here.

\acknowledgments  We thank Jordi   Miralda-Escud\'e, Avi Loeb, Karsten
Jedamzik, Sandra Savaglio, Paul Shapiro,   Andrea Ferrara, and   Simon
White   for  comments on  an  earlier  draft  of  this paper.   We are
especially greatful to Martin  Haehnelt  for many discussions and  his
constructive  criticism on this  work.  Tom Abel acknowledges  support
from NASA grant NAG5-3923.


\vfill
\eject
\end{document}